# Carrier emission of n-type Gallium Nitride illuminated by femtosecond laser pulses

Runze Li[1,2*], Pengfei Zhu[1], Jie Chen[1], Jianming Cao[1,3], Peter M. Rentzepis[2] and Jie Zhang[1**]

[1] *Key Laboratory for Laser Plasmas (Ministry of Education), Department of Physics and Astronomy, and Collaborative Innovation Center of Inertial Fusion Sciences and Applications (CICIFSA), Shanghai Jiao Tong University, Shanghai 200240, China*

[2] *Department of Electrical and Computer Engineering, Texas A&M University, College Station, TX 77843, USA*

[3] *Physics Department and National High Magnetic Field Laboratory, Florida State University, Tallahassee, Florida 32310, USA*

\* Email: Runze@ece.tamu.edu; \* \*Email: jzhang1@sjtu.edu.cn

**Abstract**

The carrier emission efficiency of light emitting diodes is of fundamental importance for many technological applications, including the performance of GaN and other semiconductor photocathodes. We have measured the evolution of the emitted carriers and the associated transient electric field after femtosecond laser excitation of n-type GaN single crystals. These processes were studied using sub-picosecond, ultrashort, electron pulses and explained by means of a "three-layer" analytical model. We find that, for pump laser intensities on the order of $10^{11}$ W/cm$^2$, the electrons that escaped from the crystal surface have a charge of ~2.7 pC and a velocity of ~1.8 μm/ps. The associated transient electrical field evolves at intervals ranging from picoseconds to nanoseconds. These results provide a dynamic perspective on the photoemission properties of semiconductor photocathodes.

**Introduction**

GaN is an important semiconductor owing to its large 3.4 eV direct band gap, which makes it extremely attractive for modern industrial applications, such as ultra-violet laser diodes [1-4] and blue light emitting devices (LEDs) [5, 6], which have culminated in the creation of white LEDs [7, 8] that are revolutionizing the industry owing to their wide emission spectra and low energy consumption [9]. In addition to those industrial applications, GaN is also a potential photocathode material because of its high quantum efficiency, negative electron affinity and resistance to vacuum contamination [10-12]. Traditionally, metals such as copper and magnesium have been the primary cathode materials for RF photoinjectors [13, 14]. For such applications, the fundamental near IR laser frequency was tripled to match the work function of metallic cathodes, where ultrashort electron pulses were generated through single photoemission and subsequently accelerated by DC or RF electrical fields. Recently, it has been shown that [15, 16], due to enhanced multi-photoemission, induced by intense femtosecond laser pulse illumination, metallic cathodes can deliver electron pulses that are comparable to those generated through single-photon process while eliminating the conversion of low-frequency photons to high-frequency photons through non-linear crystals. In addition to metallic cathodes, the possibility of employing semiconductor cathodes is also worthy of reconsideration. For example, in keV ultrafast electron diffraction, the typical cathode materials used are nanometer thick metals, which are very delicate and can be easily damaged by intense laser illumination or vacuum contamination [17-20]. A robust semiconductor cathode composed of Gallium compounds is a very desirable alternative. Particularly, Gallium compounds used as cathode are capable of delivering spin polarized electrons [21-24], when interacting with circularly polarized femtosecond laser pulses, which are suitable for developing spin-resolved ultrafast electron diffraction and electron microscopes in the future [25]. Therefore, exploring the multiphoton emission of GaN is an important research area, with potentially many technological applications.

Previously, the experimental investigations on single-photoemission of Gallium compound cathodes were mainly focused on equilibrium state parameters, such as total yield and quantum efficiency [12, 26-28]. To further evaluate the quality of photo-induced electron pulses, it is necessary to understand the ultrafast time-resolved perspective of the laser-induced carrier emission properties. Carrier transport dynamics inside the sample have been widely studied with optical and THz pump-probe methods. However, owing to the lack of ultrashort probes that are sensitive to electromagnetic fields, the emission of those carriers and the associated dynamics have rarely been studied experimentally.

Utilizing Ultrafast Electron Deflections [29-33], we measured the evolution of carrier emissions originating from femtosecond laser illumination of an n-type GaN single crystal, whose direct band gap of 3.4 eV is much larger than the 1.55 eV laser excitation photons. It is interesting to note that an unexpected strong transient electric field (TEF), on the order of tens kV/m, under an fs pump pulse intensity of $10^{10}$ W/cm$^2$ and above, was always observed. Those experimental results associated with a "Three-layer" analytical model further reveal that, upon femtosecond laser excitation, on the order of $10^{10}$ W/cm$^2$, the total amount of emitted charges and collective emitting velocities were found to be proportional to the increase in laser pulse energy. For pump intensities on the order of $10^{11}$ W/cm$^2$, saturation of the total emitted charges and their emitting velocity were observed, which we believe is a consequence of the space charge limit. Those results provide a dynamical perspective on understanding the charge limit phenomena [34] of Gallium compound photocathodes [26]. It is worthy to note the difference between Ultrafast Electron Diffraction [25, 35-38] and the Ultrafast Electron Deflection method used in our studies: In ultrafast electron diffraction experiments, the electron diffraction pattern of a crystal sample is recorded at each delay time, which reveals the transient structure changes within the sample. For ultrafast electron deflection experiments, however, the probe electron pulses do not contact with the sample but travel above the sample surface at a given height. Therefore, the deflection of those probe electrons recorded at each delay time provides the evolution of the transient electrical field above the sample surface.

**Methods**

**a. Ultrafast Electron Deflection:**

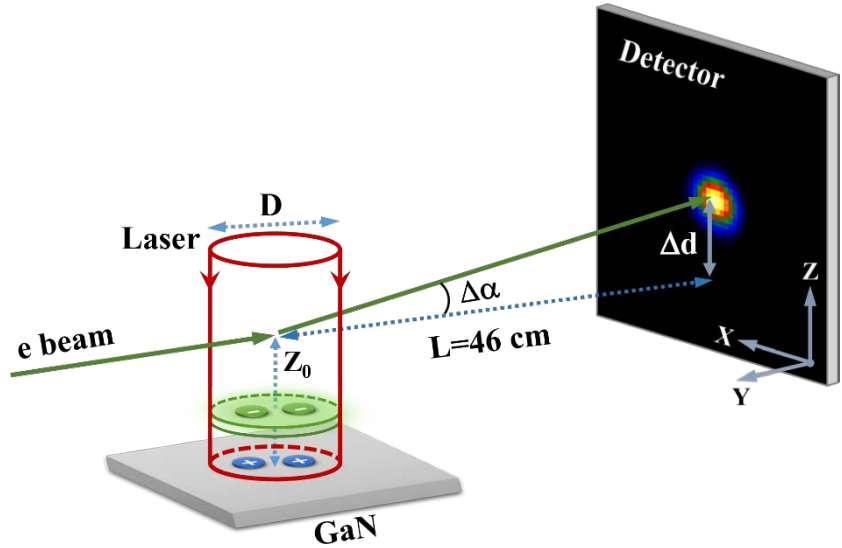

Figure 1, Experimental configuration of ultrafast electron deflection. X axis is parallel to the sample and detector surface; Y axis denotes the detector surface normal direction and is parallel to the sample surface; Z axis denotes the sample surface normal direction and is parallel to the detector surface. The pump laser pulse impinged perpendicularly onto the GaN sample with a spot diameter of D=0.8 mm ($1/e^2$). The probe electron pulse, with its centroid position $Z_0$ at 160 μm above the GaN surface, propagated parallel to the Y axis before interacting with the transient electric field area.

The experimental setup employed in this study has been described in detail elsewhere [33, 39]. In the experiments discussed here, the Ti:sapphire laser amplifier, which delivered 1 mJ, 70 fs, 800 nm laser pulses, with a repetition rate of 1 kHz, was split into two parts: one served as the pump laser pulses and impinged perpendicularly onto the sample surface; the other generated the probe electron pulses via the linear photoelectric effect, which interrogated the transient electric field above the specimen surface in a grazing angle configuration. The relative delay time between the pump laser pulse and probe electron pulse, which was determined by the difference in their traveling distance and speed to the specimen position, was precisely controlled by a linear translation stage. After being collected and amplified by an image intensifier, the 2-dimentional intensity distribution of the probe electrons, at each delay time, was recorded by a charge-coupled device (CCD) with 1.5-s exposure time. Each 2-dimetional intensity distribution was then integrated over the horizontal and vertical directions to obtain two 1-dimentional intensity distributions, which were fitted with Gaussian functions to obtain the centroid position of probe electrons.

Because the probe electrons measured the transient electric field in a grazing configuration and the pump laser employed a perpendicular incidence, the deflection of the probe electron centroid took into account only the deflection along the normal direction of the sample surface (Z axis), Figure 1. Time zero in the current studies was defined as the onset of the observable change of the probe electron centroid. After determining the time zero, the relative change of the probe electron centroid along the Z axis, $\Delta d$, was obtained by subtracting its averaged centroid position before time zero. The deflection distance on the detector ($\Delta d$) was further converted into deflection angle through $\Delta \alpha = \Delta d / L$, where L=46 cm is the distance between the sample position and the detector. The negative or positive sign of the deflection angle represents the direction of the transient electric field at the probe electron centroid position along the negative or positive direction of the Z axis, respectively. The deflection angle obtained at each delay time was averaged over 20 sets of independent measurements to improve the signal-to-noise ratio.

The pump laser intensities varied from $3.9 \times 10^{10}$ W/cm$^2$ to $4.4 \times 10^{11}$ W/cm$^2$ ( 2.7 ~ 31.1 mJ/cm$^2$), which are far below the 5.4 J/cm$^2$ damage threshold of GaN [40]. The 55 keV probe electron beam, which contained ~100 electrons, sustained a diameter of ~ 200 μm at the sample position with its centroid at ~ 160 μm above the sample surface. The sample used in our studies was a 30 μm thick, extrinsically undoped, n-type alpha <0001> GaN single crystal grown on a sapphire substrate. It is purchased from MTI Corp (Item# GN050505S-ALC30) and grown by hydride vapor phase epitaxy (HVPE) based method.

**b. "Three-layer" Model**

The deflection angle of the probe electrons represents the transient electric field at 160 μm above the GaN surface where the probe electron centroid is located. The relation between the deflection angle $\Delta\alpha_z$ and the averaged TEF at the centroid position $\bar{E}_z$ is given by:

$$\Delta\alpha_z = \frac{\Delta V_z}{V_e} = \frac{q\bar{E}_z t}{mV_e} = \frac{q\bar{E}_z t}{m_e V_e / \sqrt{1 - V_e^2/c^2}} \tag{1}$$

where $q$ and $m_e$ are the electron charge and rest mass, respectively; $c$ is the speed of light; $m$ and $V_e$ are the relativistic mass and velocity of the 55 keV probe electrons, respectively; $\Delta V_z$ is the velocity change of the probe electrons along the normal direction of the sample surface; $t$ is the interacting time between the probe electrons and the transient electric field. In our experiments, t=6 ps, which is determined by the travelling time of the 55 keV sub-picosecond probe electron pulses within the 0.8 mm pump beam diameter. Substituting the positive maximum deflection angles shown in Figure 2 into Eq.(1), we find that the maximum TEFs, at 160 μm above the GaN surface varies from 5.4 kV/m to 84.5 kV/m with the pump intensities varying from

3.9 to 44.4×10$^{10}$W/cm$^2$.

The TEFs are contributed by the electrons that escaped from the GaN sample irradiated with femtosecond laser pulses. The evolution of the escaped electrons and their interaction with the positive surface ion layer is a complicate many-body interaction process, which requires complex numerical simulations to describe the evolution of each individual charged particle. Aiming at describing the collective motion of the charges and their contribution to the TEF above the sample surface, we utilized an analytic "three-layer" model, which quantitatively provides the key parameters of the charges that determined the TEFs, such as the amount of charges emitted and the velocity of those emitted charges. This model has proven to be effective and has been discussed in detail previously [33].

In the "three-layer" model, charges that contribute to the TEFs above the sample surface are classified into three categories (three layers): (a) the positive surface ions that remain on the sample surface; (b) the emitted electrons that finally fall back onto the sample (fallen-back electrons); (c) the emitted electrons that effectively escaped from the sample (effectively emitted electrons). The effectively emitted electrons move away from the sample surface with a center-of-mass velocity $v_0$, which is also the initial emitting velocity of the electrons at time zero. The fallen-back electrons decelerate from the same initial center-of-mass velocity to zero and then reflow to the sample surface where they partially neutralize the surface ions. For the distribution function that describes the emitted charges, $\rho(z,t)$, the transient electric filed sensed by the probe electrons at the centroid position $z_0$ is given by [33]:

$$E_z(z_0, t) = \frac{\sigma_0}{2\varepsilon_0} \cdot \left\{ \left[ 1 - \frac{z_0}{\sqrt{z_0^2 + (D/2)^2}} \right] \cdot \left[ 1 - \int_{-\infty}^{0} \rho(z,t) dz \right] \right.$$

$$- \int_0^{z_0} \rho(z,t) \cdot \left[ 1 - \frac{z_0 - z}{\sqrt{(z_0 - z)^2 + (D/2 + v_w t)^2}} \right] dz \quad (2)$$

$$\left. + \int_{z_0}^{+\infty} \rho(z,t) \cdot \left[ 1 - \frac{z - z_0}{\sqrt{(z_0 - z)^2 + (D/2 + v_w t)^2}} \right] dz \right\}$$

Together with Eq. (1), the electron deflection angles under different pump intensities are fitted and depicted in Figure 2. It is worth mentioning that, the diameter of the surface ion layer, $D$, is assumed to be unchanged for the semiconductor samples studied here. This is because the carrier mobility inside semiconductors is degraded compared to that inside metals. Therefore, those carriers have negligible contribution to the neutralization of surface ions at the picosecond time scale. The emitted electrons, however, are assumed to expand with a velocity of $v_w$ along the transversal direction. $\sigma_0$ and $\varepsilon_0$ are the area charge density of the initially emitted electrons at time zero and the electrical permittivity of vacuum, respectively. The key fitting parameters that determined the observed evolutions of TEFs under various pump intensities are the initial emitting velocity of the electrons $v_0$, the total emitted charges $Q_t$ and the effectively emitted charges $Q_e$.

**Results and discussion:**

**a. Temporal evolution of probe electron deflection angles**

The time-dependent evolution of the probe electron centroid, represented by its deflection angle, is depicted in Figure 2 for pump intensities varying from 3.9 to $44.4 \times 10^{10}$ W/cm$^2$. The evolutions of deflection angles shown in Figure 2 have been vertically shifted by 0.1 mrad with respect to one another pump intensity for a clearer view of the results. For each pump intensity, the temporal behavior of the electron deflection angle consists of three steps: (1) The deflection angle reaches its negative maximum (the nadir of each curve) within 50 ps; (2) The deflection angle increases from its negative maximum to zero deflection and then reached its positive maximum deflection (the apex of each curve) in approximately 150 ps; (3) the deflection angle follows a nanosecond restoring path toward its position before time zero. Despite those universal behaviors, the magnitude and characteristic time of those deflection angles have different trends. With pump intensities below $1.3 \times 10^{11}$ W/cm$^2$, it is concluded from the data of Figure 2 that: (1) the negative and positive maximum deflection angles increased as a function of the pump intensity; (2) the time constant of the negative and positive maximum deflections also decreased as a function of the pump intensity. For pump intensities higher than $1.3 \times 10^{11}$ W/cm$^2$, the positive maximum deflection increased slightly, while negative maximum deflection and the time constant of the deflection hardly changed. The pump intensity dependent negative and positive deflection maxima are illustrated in Figures 3 and 4, respectively. We also performed the similar measurements on sapphire substrates and we did not observe electron defections at even the highest pump fluency used for the GaN sample. This suggests strongly that the electron deflection data of GaN is not affected by the sapphire substrate. In addition, previous studies have shown that, to obtain an observable electron deflection signal from sapphire, a pump fluency of 2.2 J/cm$^2$ is required, which is ~70 times higher than the largest pump fluency used in our experiments [31].

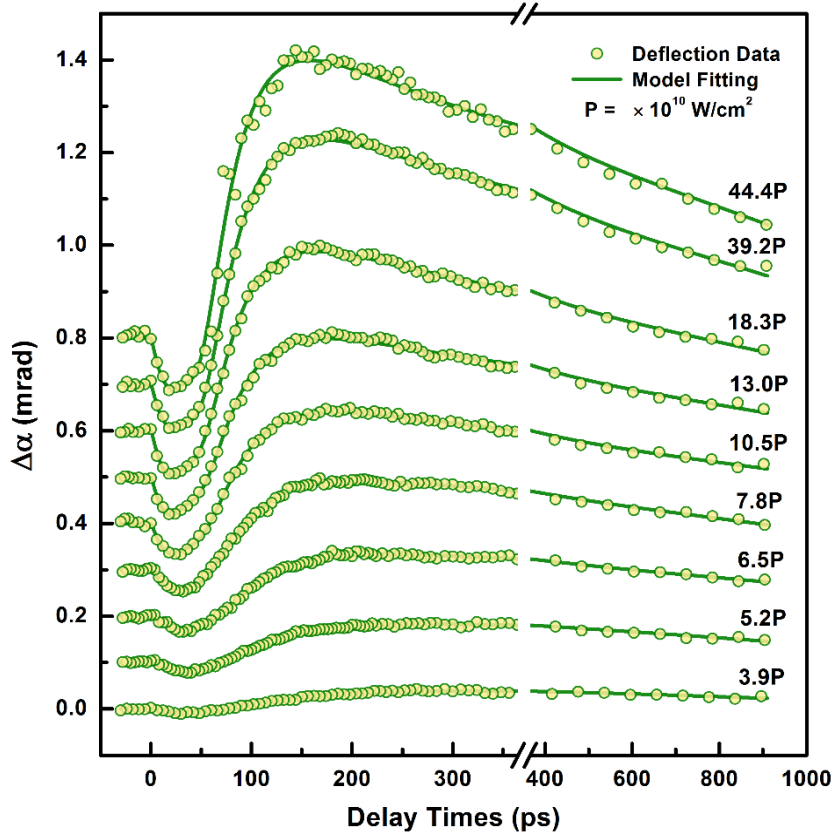

Figure 2: Time dependent evolution of the probe electron deflection angles under different pump intensities. The nadir of each curve, which corresponds to the largest negative deflection, shows that the probe electron centroid is deflected, fast, away from the sample surface and the direction of TEF at the probe electron centroid position is along the negative direction of Z axis. The apex of each curve, which corresponds to the maximum positive deflection, shows that the probe electron centroid is closest to the sample surface and the direction of TEF at the probe electron centroid position is along the positive direction of Z axis. The deflection angles under different pump fluencies are vertically shifted by 0.1 mrad with respect to one another to provide a clear view of each individual curve.

## b. The charges emitted from the n-type GaN

The deflection angles of the probe electrons shown in Figure 2 were used in connection with the "three-layer" model, discussed in the previous section. They provided insight into the total emitted electron charges $Q_t$, the effectively emitted electron charges $Q_e$, and the initial velocity of the emitted electron charges $v_0$, which are depicted in Figure 3, 4 and 5, respectively.

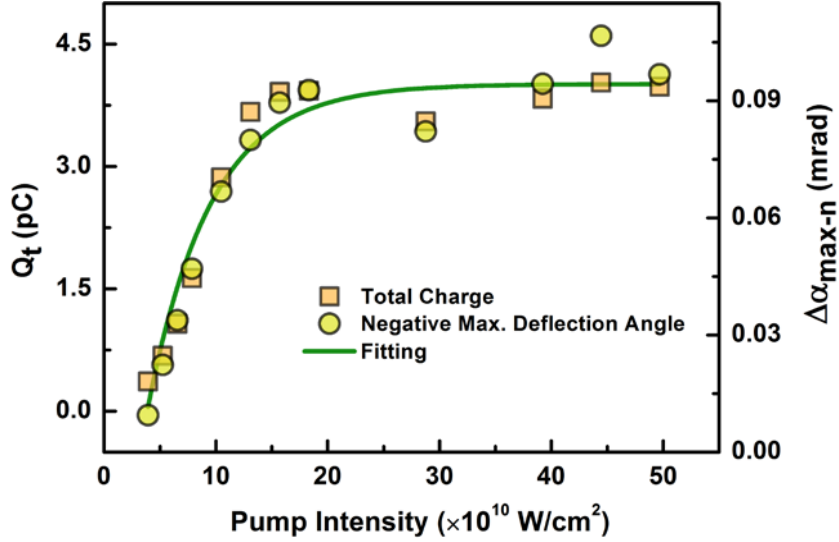

Figure 3: Total emitted electron charges and the absolute value of negative maximum deflection angles under different pump intensities. The negative maximum deflection angles are the nadirs shown in Figure 2.

The total charge of the emitted electrons at time zero is shown in Figure 3, together with the maximum deflection angles along the negative direction of the TEF. It is found that, the total electron charge and the negative maximum deflection angle, which are extracted from Figure 2, follow a similar trend with respect to the pump intensities. This indicates that, the negative maximum deflection of the probe electrons is dominated by the total emitted charges. For the first few tens of picoseconds after laser excitation, the initially emitted electrons are well below the centroid of the probe electrons and are moving toward it. Therefore, the TEF strength at the centroid position increases accordingly. However, accompanying the partial fallen back of the initially emitted electrons, the deflection angles decrease from the largest negative deflection (nadir) toward zero deflection. In addition, some of the emitted electrons eventually bypass the centroid position of the probe electrons and reverse the direction of the TEFs at that position. As a consequence, the deflection of the probe electrons starts to increase towards the positive maximum (apex). We also find that the total charges, shown in Figure 3, saturate at pump intensities higher than $\sim 1.3 \times 10^{11}$ W/cm$^2$.

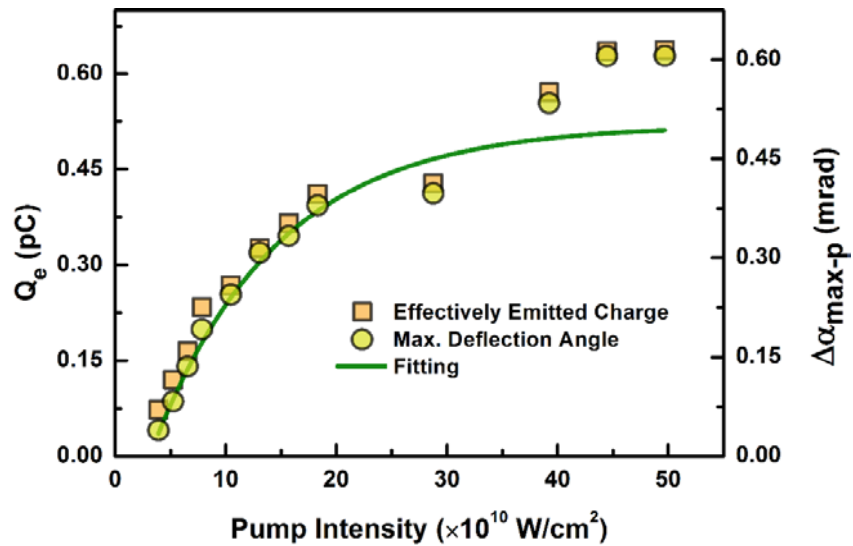

Figure 4: Effectively emitted electron charges and the maximum deflection angle of probe electrons under different pump intensities. The maximum deflection angles of probe electrons are the apexes shown in Figure 2.

The effectively emitted electrons account for 10%~20% of the total emitted charges in accord with the "three-layer" model. Those effectively emitted charges and the maximum deflection angle of the probe electrons under different pump intensities are shown in Figure 4. It is important to note that both of them have also a similar trend with respect to the pump intensity. This indicate that the maximum deflection of the probe electrons is dominated by the amount of effectively emitted electrons. After ~ 150 ps, the fallen-back electrons partially neutralize the ion layer on the GaN surface.

The transient electric field at the centroid position of the probe electrons is solely determined by the effectively emitted electrons and the remaining surface ion layer. However, the TEFs are decreased because effective emitted charges are moving away and its charge density is decreased. Therefore, the deflection of the probe electrons follow a restoring process as shown in Figure 2. Similar to the total emitted charges, the effectively emitted charges also exhibit a saturation trend for pump intensities higher than ~$1.3 \times 10^{11}$ W/cm$^2$. However, the experimentally observed saturation deviates by ~25% with respect to the saturation value predicted by the exponential fitting used in Figure 4. This is due to the nonlinear evolution of the emitted charges, which are expected to be very sensitive to fluctuations in the initial conditions of the emitted charges'.

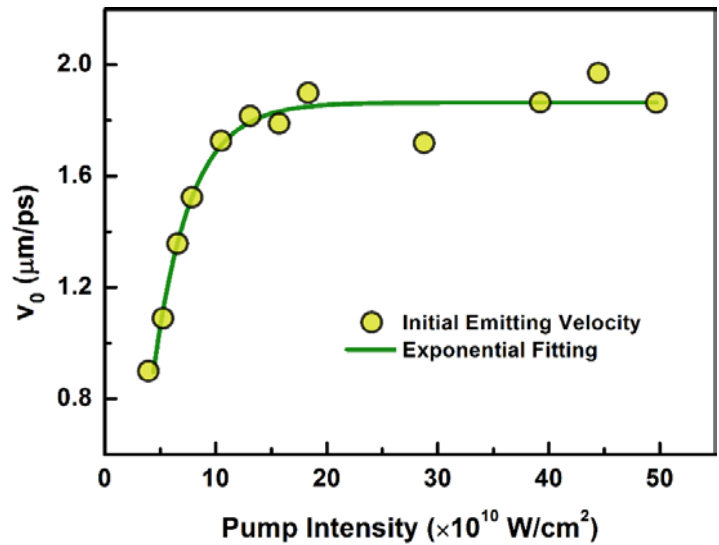

Figure 5: Intensity dependence of the initial emitting velocity of the electrons escaped from the GaN surface. This velocity represents the center-of-mass velocity of the initially emitted electrons at time zero.

The initial emitting velocity $v_0$, which represents the center-of-mass velocity of the escaped electrons at time zero, is shown in Figure 5. Similar to the total emitted charges and effectively emitted charges, the initial emitting velocity increases almost linearly with the pump intensity growth, but remains almost constant for pump intensities larger than ~$1.3 \times 10^{11}$ W/cm$^2$. The emitting velocity mainly determined the characteristic time of the probe electron deflections. At higher emitting velocities, it takes shorter time for the emitted electrons to reach the centroid position of the probe electrons and change the direction of TEF at that position. Consequently, the time constant of probe electrons to reach the negative maximum deflection is shorter, which agrees with the time shifting of the nadirs shown in Figure 2. Meanwhile, the time required for the probing electrons to reach the positive maximum deflection is decreased. The fluctuations within the saturation fluencies shown in Figure 3, 4, and 5 are mainly due to the nonlinear evolutions that are sensitive to the initial conditions of the emitted charges.

## c. Photoemission properties of the n-type GaN

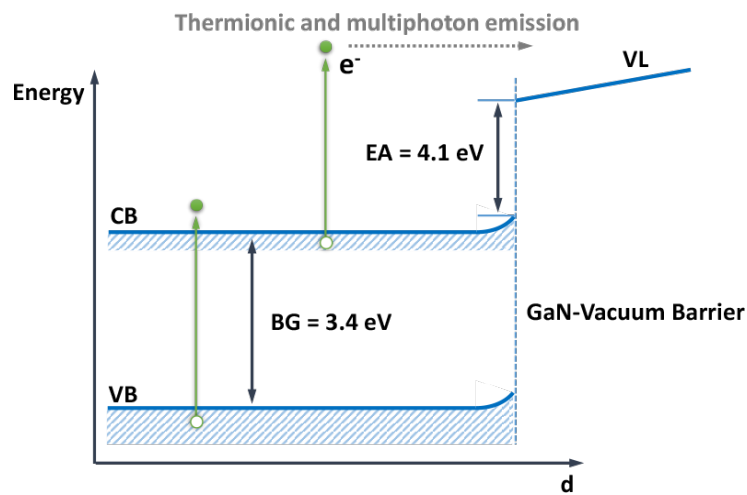

Figure 6: Schematic illustration of the GaN band structure and possible excitation channels because of femtosecond laser irradiation. VL, CB, VB, EA and BG represent Vacuum Level, Conduction Band, Valance Band, Electron Affinity and Band Gap, respectively.

A simplified band structure of extrinsically undoped, n-type GaN is illustrated in Figure 6. Upon femtosecond laser pulse irradiation, the photon energy is rapidly deposited into the free carriers, which are the conduction band electrons in the case of n-type GaN. Part of the excited electrons overcome the vacuum level energy, which is mainly defined by the electron affinity of 4.1 eV [28, 41], and escape from the sample surface to vacuum through thermionic or three-photon emission. The valance band electrons may also escape from the GaN surface through five-photon process, which is typically orders of magnitude smaller than those emitted through thermionic or three-photon emissions. The total yield of the photoemission is described by the Fowler-DuBridge theory [42-46], which integrates the contributions from both thermal and multiphoton mechanisms. According to this theory, the thermal yield increases linearly as a function of the laser intensity [33], while the n-th order photoemission yield is proportional to $I^n$, where $I$ is the pump laser intensity [15]. Based on the data shown in Figure 3, a linear fitting in the log-log plot of the total charges, for pump intensities below $1.3 \times 10^{11}$ W/cm$^2$ indicate a slope of 1.96, which corresponds to $Q_t \propto I^{1.96}$. This indicates that the emitted electrons through the three-photon emission take a considerate portion of the total emitted electrons.

The carrier concentration of the n-type GaN crystal used in this study is $n_e \approx 1.0 \times 10^{17}$ cm$^{-3}$ at room temperature, which is measured by the Hall-effect and agrees with previous results [47]. In the widely accepted "three-step" model that describes photoemission processes, only the free carriers within the mean free path of the hot electrons [48] can possibly diffuse to the sample surface, namely overcome the vacuum energy level and be eventually emitted. For the GaN specimen, used here, the hot electron mean free path is considered as $L_d$ =10 nm [49]. Taking into account the diameter of the laser excitation area, $D$, the total number of electrons in the conduction band given by $N_e = n_e L_d \pi (D/2)^2$, which equals to $5.0 \times 10^8$ electrons and corresponds to a charge of 80 pC. In the experiments presented here, the saturated emission of the photoelectrons given in Figure 3 corresponds to a total charge of 3.8 pC, which accounts for 5% of the conduction band electrons. Therefore, such saturation is not due to the limited number of conduction electrons. The saturation observed may be explained by: (i) Space charge limit. The electrons that escaped from the sample surface at an earlier time formed the TEF which suppressed the emission of electrons at a later time. This effect has been observed in previous photoelectron gun studies [50, 51]. To overcome the effect of space charge limit, an applied extraction field, that is much large than the TEF, is typically employed. (ii) Modification of GaN-Vacuum barrier. Previous studies of photocathodes based on Gallium compounds have revealed that, even if a sufficient high extraction field is applied to overcome the effect of space charge limit, the photoelectrons may remain saturated [26]. It is believed that the emitted electrons establish a time-evolving photo-voltage on the sample surface and change the surface state, which eventually increases the vacuum level and limits the number of emitted electrons [27]. In addition to the saturation of emitted charges, it is also found that, for pump intensities above $\sim 1.3 \times 10^{11}$ W/cm$^2$, the emitting velocity remains unchanged at $\sim$ 1.8 μm/ps, as shown in Figure 5. Such a dynamical equilibrium for the interaction between photons and GaN, at higher illumination intensities, indicates a stable photoelectron emission state. Therefore, GaN irradiated with high laser intensities might be used as a high quality femtosecond electron sources in terms of stable quantum yield and smaller energy dispersion.

Comparing with our previous studies of Al [33], it is found that the total emitted charges and effectively emitted charges from the GaN film are always smaller than those of Al under similar pump laser fluencies. This is probably due to the larger number of free electrons in metals than semiconductors. Meanwhile, the slope of the fallen-back ratio versus pump fluency is essentially the same for both Al and GaN, indicating that the nonlinear evolution of the emitted charges follows the trend that with more electrons escaping from the sample, the stronger Coulomb repulsion forces more electrons fallen back into the sample. However, it is worthy to mention that, the saturation of the emitted charges was not observed in the studies of Al samples. This is because the 25 nm thick freestanding Al samples are too fragile to be pumped with high laser fluencies. Therefore, the pump fluencies used in the Al experiments may be insufficient to excite enough photoelectrons to reach the space charge limit [50, 51].

Despite the photoemission properties of GaN revealed in this study, further experiments are required to identify whether the mechanism of charge saturation is due to space-charge effect or the modification of the GaN-Vacuum barrier. One possible method that may determine the mechanism is to measure the photoemission of GaN specimens that have different electron affinity energies. The dynamical properties introduced here imply that a saturation region exists for femtosecond laser pulse-GaN interaction, in which stable electron pulses could be realized whose total charge and initial emitting velocity are insensitive to the fluctuations in laser pulse energy. Previous studies of Mg-doped GaN cathodes with a negative electron affinity revealed that the quantum efficiency of single-photoemission could reach 70% [11]. The QE of the GaN sample in our experiment is about five orders of magnitude smaller than those of previous reports. This is mainly because the efficiency of multiphoton emission is typically orders of magnitude smaller than single-photoemission. However, it is worthy to note that the efficiency of multiphoton process increases exponentially with the width of the excitation pulse, therefore, shorter fs pulses will be more efficient. Meanwhile, it has been demonstrated that the multiphoton emission efficiency of Gallium compounds could be improved by Plasmon excitation [24]. Although our electron yield is on the order of pC, it is more than enough for applications such as keV ultrafast electron diffraction where the space charge effect is generally minimized through decreasing electron numbers to maintain a femtosecond pulse width [52]. The studies presented here also provide means for diagnosing the properties of photocathodes from a dynamics perspective, such as the evolution of surface electric fields and charge emitting properties. Additional diagnostic dimension connecting the electron pulses would help to understand interactions between those TEFs and the acceleration field applied to a cathode, which may further facilitate the generation of high quality electron beams from GaN.

**Conclusion:**

The photoemission properties of a semiconductor photocathode material, GaN, has been studied experimentally using ultrafast electron deflection and explained using the "three-layer" model. Upon illumination of femtosecond laser pulses, the n-type GaN single crystal is found to emit electron pulses with pC charges through thermionic and three-photon emission. The saturation of the total emitted charges and initial emitting velocity implies that a stable electron bunch may be achieved at pump intensities higher than $1.3\times10^{11}$ W/cm$^2$. In addition, to reveal the photoemission properties of GaN from a dynamics perspective, our experimental studies also provide a means for monitoring, directly, the transient electric fields, built up at a photocathode surface.

**Acknowledgements:**

This research was supported in part by the National Natural Science Foundation of China (Grant Nos. 11675106 and 11421064), the National Basic Research Program of China (Grant No. 2013CBA01500), the Welch Foundation (Grant Number: 1501928) and Texas A&M University TEES.